# Develop machine learning based predictive models for engineering protein solubility

Xi Han[1], Xiaonan Wang[1,*], Kang Zhou[1,*]

[1]Department of Chemical and Biomolecular Engineering, National University of Singapore, Singapore, 117585.

*To whom correspondence should be addressed.

## Abstract

**Motivation:** Protein activity is a significant characteristic for recombinant proteins which can be used as biocatalysts. High activity of proteins reduces the cost of biocatalysts. A model that can predict protein activity from amino acid sequence is highly desired, as it aids experimental improvement of proteins. However, only limited data for protein activity are currently available, which prevents the development of such models. Since protein activity and solubility are correlated for some proteins, the publicly available solubility dataset may be adopted to develop models that can predict protein solubility from sequence. The models could serve as a tool to indirectly predict protein activity from sequence. In literature, predicting protein solubility from sequence has been intensively explored, but the predicted solubility represented in binary values from all the developed models was not suitable for guiding experimental designs to improve protein solubility. Here we propose new machine learning models for improving protein solubility *in vivo*.

**Results:** We first implemented a novel approach that predicted protein solubility in continuous numerical values instead of binary ones. After combing it with various machine learning algorithms, we achieved a prediction accuracy of 76.28% when Support Vector Machine (SVM) algorithm was used. Continuous values of solubility are more meaningful in protein engineering, as they enable researchers to choose proteins with higher predicted solubility for experimental validation, while binary values fail to distinguish proteins with the same value – there are only two possible values so many proteins have the same one.

**Availability:** We present the machine learning workflow as a series of IPython notebooks hosted on GitHub (https://github.com/xiaomizhou616/protein_solubility). The workflow can be used as a template for analysis of other expression and solubility datasets.

**Contact:** kang.zhou@nus.edu.sg, chewxia@nus.edu.sg

**Supplementary information:** Supplementary data are available on GitHub (https://github.com/xiaomizhou616/protein_solubility).

## 1 Introduction

*Escherichia coli* is a bacterium commonly used in genetic engineering to express recombinant proteins (Chan, et al., 2010), which is a pivotal process in biotechnology. Producing some proteins in *E. coli* was, however, inefficient in industrial setting, resulting from low activity of the recombinant proteins. As biocatalysts and therapeutic agents, proteins with high activity have lower unit production cost. Developing biocatalysts with high activity has thus become an important goal for further development of many biocatalytic processes.

Some experimental strategies *in vivo* can improve the expression of recombinant proteins, such as using suitable promoters, optimizing codon usage, or changing culture media, temperature and/or other culture conditions (Idicula‐Thomas and Balaji, 2005; Magnan, et al., 2009). Such empirical optimizations have, however, been time-consuming and expensive. Moreover, experiments often fail due to opaque reasons. A generic solu-tion is highly desired for enhancing the heterologous protein overexpression and may be ultimately provided by using a computational model that can predict activity of any enzyme accurately from its amino acid sequence and other input information. Developing such model would require a large dataset which contains at least two columns, protein sequence and activity. No such dataset is currently available, because conventional protein engineering was driven by chasing high activity, which has provided very little information on protein sequence. For instance, most proteins with low or intermediate activity values were discarded without their sequence being revealed. It is currently prohibitively expensive to generate a complete, large dataset for protein activity, due to technical limitations.

A compromised solution is to predict protein activity by using solubility of the protein as a proxy, because (1) activity and solubility are correlated for some proteins, and (2) there is a relatively large dataset available for solubility as solubility data from different proteins can be pooled together. Until now, a number of machine learning predictors have been developed to address the interconnection between protein solubility and amino acid sequence. The prediction for protein solubility from amino acid sequence was first proposed by Wilkinson and Harrison (Wilkinson and Harrison, 1991). A simple regression method was used by them to achieve high accuracy (0.88) by using several features of 81 protein sequences, such as hydrophilicity, molecular weight and averaged surface charge. The accuracy was improved by logistic regression to 0.94 based on a small database with 212 proteins (Diaz, et al., 2010). SVM is a core machine learning (ML) classification method that has achieved competitive performance in this field. With SVM, Idicula-Thomas analysed 192 proteins with an unbalanced correlation score filter, which achieved an accuracy of 0.74 by using physicochemical properties, mono-peptide frequencies, dipeptide frequencies and reduced alphabet set as features (Idicula-Thomas, et al., 2005). These datasets are, however, not large enough to generate correlations that can be widely applied (Idicula-Thomas and Balaji, 2005). Molecular weight, isometric point and amino acid composition were used as inputs to build a SVM model by Niwa et al. (Niwa, et al., 2009), who for the first time generated and studied a dataset that contained more than 1000 entries. Prediction of protein solubility was subsequently conducted with SVM based on databases with 2159 proteins (Agostini, et al., 2012) and 5692 proteins (Xiaohui, et al., 2014). Besides, decision tree (Christendat, et al., 2000; Goh, et al., 2004; Rumelhart, et al., 1985), Naïve Bayes (Smialowski, et al., 2006), random forest (Fang and Fang, 2013; Hirose, et al., 2011), gradient boosting machine (Rawi, et al., 2017) and some computational methods combining multiple machine learning models have been utilized to predict if a protein is soluble or not by using their primary sequence. In addition, several software and web servers have been developed for protein solubility prediction, including ESPRESSO (Hirose and Noguchi, 2013), Pros (Hirose and Noguchi, 2013), SCM (Huang, et al., 2012), PROSOII (Smialowski, et al., 2012), SOLpro (Magnan, et al., 2009) and PROSO (Smialowski, et al., 2006). . Some of the predictors available online are introduced below. ESPRESSO selects features by using Student's t-test filter and trains a model by SVM and another sequence pattern recognition based method. Pros implements random forest with Student's t-test as filter and random forest as wrapper. PROSOII is a two-layer model using a Parzen window in the first layer and logistic regression in both layers.

Various databases were utilized in the previous studies, whereas eSol database (Niwa, et al., 2009) is a unique one, which has a relatively large size and continuous values of solubility, ranging from 0 to 1. Comparison of model performances using this database is summarized in Table 1. All the ML models developed based on this database, however, only predicted whether a protein was soluble or insoluble (outputting binary values), which was not useful in guiding protein engineering, because even a substantial increase in solubility may not turn the status from 'insoluble' to 'soluble' and thus be falsefully viewed as useless by the model. Moreover, it is challenging to select several proteins with highest solubility to conduct experiments because there are too many proteins with the same value 1 of solubility in a large dataset.

In the present study, we built models that can predict protein solubility in numerical values that span from 0 to 1, which would allow fair evaluation of a proposed mutation to a protein even when the resulting improvement is small. We compared various model-building techniques which use this new output format, and found SVM performed the best, achieving an accuracy of 76.28%.

## 2 Methods

### 2.1 Protein database

All the proteins used in the study were downloaded from eSol database (Niwa, et al., 2009). A brief summary of how solubility values in eSol database were obtained is provided here. Proteins were produced *in vitro* by using cell-free protein expression technology and plasmids carrying Open Reading Frame (ORF) from complete *E. coli* ORF library (ASKA library) (Kitagawa, et al., 2005). Synthesized proteins were fractionated into soluble and insoluble fractions by using centrifugation, and the proteins in both fractions were quantified by using SDS-PAGE. Solubility was defined as the ratio of protein quantity of the supernatant to the total protein quantity. We used all the entries from the database, unless the entries contain no sequence information, or poorly determined sequence (containing N instead of A, T, C, or G, or multiple stop-codons). We excluded 26 entries in total, and the size of the dataset used was 3147.

### 2.2 Features

Different features were extracted from sequences of proteins by using protr package (Xiao, et al., 2014) within R software, which generates various numerical representation schemes. In our study, the description of features is listed in Table 2 according to different descriptors in protr package.

### 2.3 Training flowsheet

Data pre-processing is a very important part in data mining, which transfers raw data into a well-organized dataset that can be taken as inputs in model training. Only sequence and protein solubility were retrieved from the original dataset. Protein sequence was then transformed into numeric values (features) by using an algorithm from Table 2. Finally, data normalization was conducted by making numerical values of each column in the range of 0-1.

Training machine learning models for protein solubility prediction included several steps (as shown in Fig. 3). Seventy five percent of the original dataset (2364 proteins) were used in GANs model training and 90 percent of the original dataset (2833 proteins) were used in other machine learning model training. The remaining dataset were validation data which can be used to evaluate the performance of each model. After model training, all the evaluation metrics were calculated with 10-fold cross validation to compare the performance of different machine learning models.

### 2.4 Machine learning models

Seven supervised machine learning algorithms were applied to our dataset: logistic regression, decision tree, SVM, Naïve Bayes, conditional random forest (cforest), XGboost and artificial neural networks (ANNs). For binary values of solubility, classification algorithms were used and for continuous values of solubility, regression algorithms were applied.

### 2.5 Tuning SVM

Among the machine learning models used to predict the protein solubility, SVM achieved the best performance, so SVM was chosen for further optimization to achieve higher prediction accuracy. The parameters of SVM were tuned (Package e1071 in R software was used to implement SVM model). Different kernels in SVM represent different functions. Another three parameters tuned were cost, gamma and epsilon. Cost is the regularization parameter that controls the trade-off between achieving a low training error and a low testing error. Epsilon is a margin of tolerance where no penalty is given to errors. Gamma indicates the threshold to determine if two points are considered to be similar.

**Table 1**. Performance of published methods based on eSol database

| Method | Size | Accuracy | Reference |
|---|---|---|---|
| Random forest | Total: 1918 | 0.84 | (Fang and Fang, 2013) |
| | Soluble: 886 | | |
| | Insoluble: 1032 | | |
| 1. Support vector machine | Total: 1625 | 0.81[a] | (Samak, et al., 2012) |
| 2. Random forest | Soluble: 843 | | |
| 3. Conditional inference trees | Insoluble: 782 | | |
| 4. Rule ensemble | | | |
| Decision tree | Size: 1625 | 0.75 | (Stiglic, et al., 2012) |
| | Soluble: 843 | | |
| | Insoluble: 782 | | |
| Support vector machine | Size: 2159 | N/A | (Agostini, et al., 2012) |
| | Soluble: 1081 | | |
| | Insoluble: 1078 | | |
| Support vector machine | Size: 3173 | 0.80 | (Niwa, et al., 2009) |
| | Soluble: >0.7 | | |
| | Insoluble: <0.3[b] | | |

[a] Value estimated from the figure in the reference

[b] Proteins with solubility higher than 0.7 are labelled as soluble and lower than 0.3 are labelled as insoluble

**Table 2.** Description of different descriptors in protr package of R software

| Descriptors | Description |
|---|---|
| extractAAC | composition of each amino acid |
| extractPAAC | hydrophobicity, hydrophilicity and side chain mass |
| extractAPAAC | hydrophobicity, hydrophilicity and sequence order |
| extractMoreauBroto | normalized average hydrophobicity scales, average flexibility indices, polarizability parameter, free energy of solution in water |
| extractMoran | normalized average hydrophobicity scales, average flexibility indices, polarizability parameter, free energy of solution in water [c] |
| extractGeary | normalized average hydrophobicity scales, average flexibility indices, polarizability parameter, free energy of solution in water [c] |
| extractCTDC | hydrophobicity, normalized van der Waals volume, polarity, polarizability, charge, second structure, solvent accessibility |
| extractCTDT | percentage of position of attributes in extractCTDC |
| extractCTDD | distribution of attributes in extractCTDT |
| extractCTriad | protein-protein interactions including electrostatic and hydrophobic interactions |
| extractSOCN | Schneider-Wrede physicochemical distance matrix and Grantham chemical distance matrix |
| extractQSO | Schneider-Wrede physicochemical distance matrix and Grantham chemical distance matrix [d] |

[c] Different calculation methods are used for extractMoreauBroto, extractMoran and extractGeary using the same attributes listed above

[d] Different calculation methods are used for extractSOCN and extractQSO using the same attributes listed above

## 2.6 Data augmentation algorithms

Generative Adversarial Networks (GANs) are an emerging Artificial Intelligence (AI) algorithm used for data augmentation in this work. There are two main models within GANs, Generator Neural Network and Discriminator Neural Network. The generative model $G$ takes random data $z$ from probability distribution $p(z)$ as input. The discriminative model $D$ takes data generated from generative model $G$ and real data from training dataset as inputs and distinguishes whether the data are artificial or real. The training process is an adversarial game with backpropagation between two networks (Fig. 2).

In Equation (1), $D(x)$ calculates the probability that $x$ comes from the original data. $D$'s objective is to maximize the probability of labelling both the real and artificial data correctly (termed as max($V$)), whereas $G$'s objective is to minimize max ($V$), which can be expressed as min(max($V$)). Overall, it is a two-player game between $D$ and $G$.

For training GANs, discriminator is first trained for *n* epochs based on real data. Second, data artificially generated from generator is fed into discriminator networks which gave a judgement if the data was artificial. Third, generator networks were further trained with the feedbacks from discriminator to improve its data generation. The training process was iterated until the data generated from generator cannot be distinguished by the discriminator networks from the original dataset.

$$\min_{G} \max_{D} V(D, G) = E_{x \sim P_{data(x)}} \left[ \log D(x) \right] + E_{z \sim P_{z(z)}} \left[ \log(1 - D(G(z))) \right] \tag{1}$$

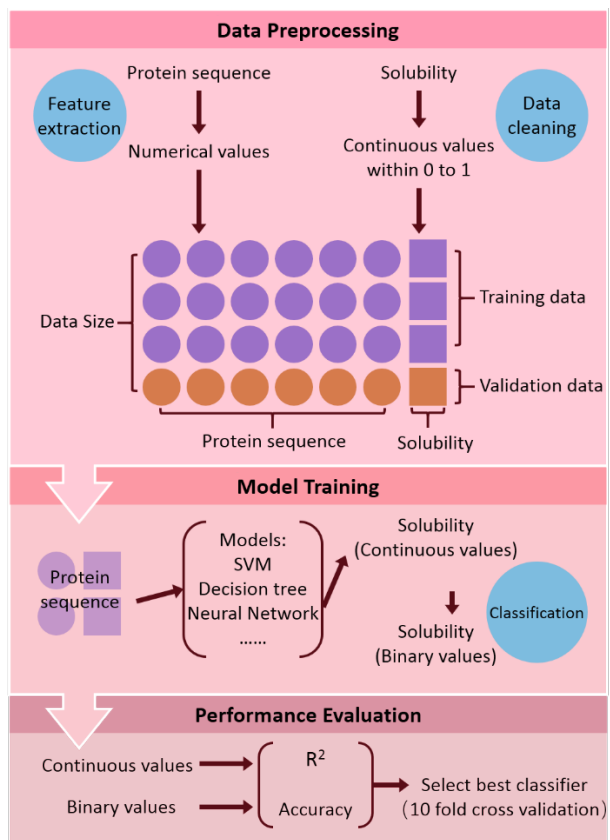

**Fig. 1. Workflow of model training used in our study**

With the development of AI algorithms for data augmentation, other algorithms based on GANs have also been developed, which enhanced the performance of GANs to some extent. CGAN is the conditional version of GANs, which generates mimic data with class labels (Mirza and Osindero, 2014). WGAN uses Wasserstein distance metric rather than Jensen-Shannon distance metric used in GANs when cross-entropy is calculated (Arjovsky, et al., 2017). The cross-entropy loss is a measurement of the ability of discriminator to distinguish real data and generated data. Wasserstein distance measures the minimum cost to make the distribution of generated data more similar to that of real data, with the critic function in the discriminator model. WCGAN is the conditional version of the Wasserstein GANs (Gulrajani, et al., 2017).

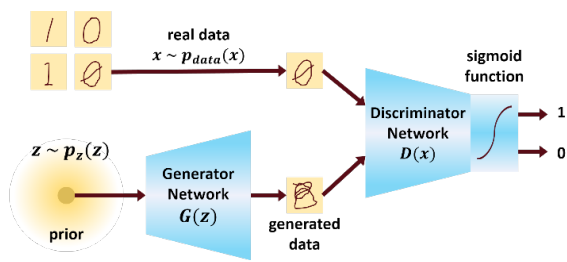

**Fig. 2. The workflow of training GANs**

## 2.7 Evaluation metrics

Several evaluation metrics were used to quantitatively evaluate the performance of different machine learning models. For comparison with prior work, we transformed continuous predicted solubility values into binary ones. Accuracy was calculated based on the binary values. The definition of TP (true positive), TN (true negative), FP (false positive) and FN (false negative) is illustrated in Table 3. How to use TP, TN, FP and FN to calculate accuracy is shown in Equation (2).

$$\text{Accuracy} = \frac{a+d}{a+b+c+d} \qquad (2)$$

**Table 3.** The definition of TP, TN, FP and FN for binary values of solubility

| Predicted solubility from models | | |
|---|---|---|
| Actual solubility from experiments | Solubility=1 | Solubility=0 |
| Solubility = 1 | a (TP: true positive) | b (FN: false negative) |
| Solubility = 0 | c (FP: false positive) | d (TN: true negative) |

We also evaluated the model performance by using coefficient of determination ($R^2$) calculated from predicted and actual solubility values. Computing $R^2$ can only be done by using continuous values our models generated.

K-fold cross-validation is a technique to partition the original dataset into $k$ equal-size parts. And each part has similar proportion of samples from different classes. In the $k$ subsamples, one subsample is taken as validation data to test the model and the other data is used as training data to train the model. The accuracy and $R^2$ are the average of all the cases. In our study, we used k-fold cross-validation, in which $k$ was set as 10.

## 3 Results

### 3.1 Develop models that output continuous numerical values

We used continuous values of protein solubility to train machine learning models, which we set to output continuous values instead of binary ones. We have trained seven distinct models and compared their performance by using two different evaluation criteria (accuracy and $R^2$, Table 4). For the purpose of calculating the accuracy, we converted the continuous values of predicted solubility into binary ones by using 0.44 (median value of the solubility data of our database) as cut-off threshold because the data above and below median exhibit same sizes. SVM showed the best prediction performance among all the machine learning models (Table 4). The tuned SVM parameters were listed in Table S1 and Fig. S1 in Supplementary Information. After plotting the correlation between the accuracy vs. cost, epsilon, or gamma, optimized values were found (1.26, 0.029 and 0.057 for cost, gamma, or epsilon, respectively).

For comparison, we also trained the models by using the conventional approach, which converted continuous actual solubility values into binary ones BEFORE model training. Though using continuous values did not always improve model prediction performance, it improved SVM model and made it to be the best one, in terms of both accuracy and $R^2$ (Table 4).

**Table 4.** Performance of different machine learning models

| Model | Accuracy | | $R^2$ |
|---|---|---|---|
| | Continuous values of solubility | Binary values of solubility | Continuous values of solubility |
| Logistic regression | 0.6372 | 0.6868 | 0.2507 |
| Decision tree | 0.6734 | 0.6750 | 0.2459 |
| SVM | 0.7538 | 0.7001 | 0.4107 |
| Naïve Bayes | 0.6674 | 0.6979 | 0.2376 |
| cforest | 0.7055 | 0.7123 | 0.3764 |
| XGboost | 0.7052 | 0.7103 | 0.3997 |
| ANNs | 0.6632 | 0.7086 | 0.2029 |

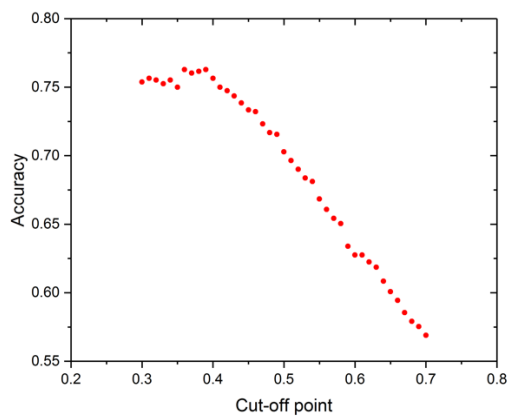

**Fig. 3.** Effect of cut-off threshold on SVM prediction accuracy

Cut-off points were used to transfer continuous values of solubility to binary values for model evaluation. How to choose the cut-off threshold substantially affects accuracy that is used to gauge model performance. In previous studies, many arbitrarily set it at 0.5. Here we did a simple optimization of the threshold with maximizing the accuracy as the objective. When the cut-off threshold was too low (< 0.3) or too high (> 0.7), the accuracy would be very high, but such accuracy was not useful because almost all the proteins would be classified either as soluble or insoluble proteins. We varied the threshold from 0.3 to 0.7 with step size of 0.01 and calculated the SVM model prediction accuracy. From Fig. 3, the highest accuracy (0.7628) was obtained when the cut-off point was 0.36 or 0.39. Therefore, 0.39 was randomly selected from those two values and was used in the following analysis. Although this optimization did not substantially improve SVM model prediction accuracy, the results showed that the threshold can substantially change the accuracy (Fig. 3). Since it is arbitrary to choose a cut-off threshold, accuracy based on the cut-off threshold is intrinsically subjective. With our new methods and the resulting continuous model output values, we call for using the coefficient of determination ($R^2$) as the model evaluation metric, which does not require any cut-off threshold. We used the $R^2$ as the sole evaluation metric in the rest of this study.

**3.2 Attempt to improve model performance by optimizing data pre-processing**

To further improve SVM model prediction performance, we evaluated all the available algorithms that convert amino acid sequence into numerical values. This step is critical as it extracts a fraction of information (usually only a small fraction, termed as features) from the sequence data, and feeds it to the model-training algorithms. Since most information in raw data was discarded here, avoiding loss of critical information by using a better algorithm may improve the model performance. It turned out that the best was the basic algorithm that simply converted sequence into amino acid composition, and that we used in prior sessions of this study (Table 5).

**Table 5**. Performance of different sequence descriptors by SVM [e]

| Descriptors | $R^2$ | Data dimension |
|---|---|---|
| extractAAC | 0.4108 | 20 |
| extractPAAC | 0.3564 | 50 |
| extractAAC+extractPAAC | 0.3719 | 70 |
| extractAPAAC | 0.3049 | 80 |
| extractMoreauBroto | 0.0806 | 240 |
| extractMoran | 0.0542 | 240 |
| extractGeary | 0.0562 | 240 |
| extractCTDC | | 21 |
| extractCTDT | 0.2484 [f] | 21 |
| extractCTDD | | 105 |
| extractCTriad | 0.0650 | 343 |
| extractSOCN | 0.1647 | 80 |
| extractQSO | 0.3224 | 100 |

[e] The description of each descriptor is shown in Table 2

[f] The $R^2$ is the result combining extractCTDC, extractCTDT and extractCTDD

## 4  Discussion

### 4.1 Predict protein solubility with continuous values

In this study, we developed a new model from amino acid sequence that can predict protein solubility as continuous values from 0 to 1, which would help improve solubility of recombinant proteins through protein engineering. Having model output in such format also enables to use coefficient of determination ($R^2$) as a metric to evaluate model performance, which does not depend on a subjective threshold to determine if a protein is soluble or insoluble. In literature, all the studies used such threshold(s) to convert continuous values of protein solubility in eSol database into binary values.

With continuous values of solubility, we will study how to use a developed model to guide experiments in the future. First, we will use statistical tools to determine the minimal performance of a model (in terms of $R^2$) required to conduct a meaningful *in silico* screening of protein mutants – the average quality of selected mutants should be substantially higher than that of the initial pool of mutants. The next step is to validate the model prediction by selecting a number of proteins with high predicted solubility and measuring their actual solubility through experiments. We plan to generate the pool of mutants for *in silico* screening by randomly introducing mutants to a model protein, However, binding of substrate/co-factor and catalytic active site may be damaged by such random mutation, which would influence the catalytic function. Therefore, mutation of residues responsible for binding substrates and the catalytic activity will be avoided. Then sensitivity analysis will be conducted to determine how many mutations should be made to improve protein solubility substantially. The model we developed in this study laid the foundation for such future works, which cannot be done by using any existing models.

## 4.2 Attempt to improve model performance by using data augmentation

Having adequate volume of data is a key to develop useful models by using machine learning. In biotechnology field, generating and collecting large amount of data is, however, time-consuming and expensive, resulting in much smaller dataset than other fields such as facial recognition. We hypothesized that data augmentation algorithms might alleviate the problem that biotechnology applications do not have sufficient data for developing machine learning models. We attempted to use GANs to improve prediction of our SVM model by generating artificial data.

In the first attempt, four versions of GANs were trained for 500 iterations to generate the artificial data, which were used together with the original data to train ML models by using SVM. This attempt, however, failed to improve the model performance, as $R^2$ was not increased (Supplementary Table S2). Over iterations, the quality of data generated by GANs improved (Supplementary Fig. S3), which showed GANs were implemented correctly. By comparing generated data with original data (Supplementary Fig. S2) and the performance of four versions of GANs (Supplementary Table S2), GANs have the best performance among the four versions of GANs. Therefore, GANs were explored for more iterations.

In the second attempt, we increased the number of iterations from 500 to 5000. Three randomly selected datasets were tested to make the results more representative (Supplementary Table S3). However, there is no improvement of prediction performance after applying data augmentation algorithms. Optimization for GANs or other data augmentation algorithms needs to be explored further.

Despite that GANs have improved ML in many applications, it has failed to build better models in this case. There may be two reasons for the undesired results. First, the parameters of GANs or the architecture of generator and discriminator were suboptimal, so the quality of the generated data was poor. To improve GANs, we plan to conduct more optimizations of GANs parameters, such as learning rate of generator and discriminator. Second, model prediction performance is strongly influenced by the quality of dataset, so the quality of the dataset we used may limit us from developing better models. In this case, GANs can be tried for a part of the original dataset only including specific groups of proteins with acceptable prediction accuracy.

## 4.3 Prediction of protein yield

The yield of biocatalysts here means the concentration of biocatalysts produced by the host cells ('titer' is used to describe it in some fields). Developing biocatalysts with high yield cuts down the unit cost of biocatalysts in pharmaceutical, chemical and food industry. Random mutagenesis for improving the yield of recombinant proteins is very time-consuming and expensive. Utilizing the data collected in experiments to guide the mutation should be very helpful. Predicting protein yield *in silico* by machine learning models has never been done. If such a model can be successfully developed, the design of better biocatalytic processes should be much easier considering useful guidance of such a model on improving protein yield. We used eSol database, which also provides protein yield in unit of μg/mL (Niwa, et al., 2009).

In yield dataset, there are some data points with very high yield values, which are considered to be outliers that will influence the modelling sig-nificantly and need to be removed. We compared two approaches to remove outliers (Supplementary Table S4) and found that the IQR approach ($R^2$=0.0759) was better than the standard deviation approach ($R^2$=0.0622), based on $R^2$, the metric used to evaluate performance of the developed models (SVM was used in this evaluation). The IQR approach was utilized in the rest of the study.

We compared SVM with other ML methods to develop models for predicting protein yield and found SVM was the best though its performance was not satisfactory – $R^2$ was only 0.1163 (Supplementary Table S5). The tuning process was conducted for the best model SVM, and optimized parameters were used for further analysis (Supplementary Table S6, Supplementary Fig. S4). We further attempted to improve the SVM model by using different descriptors and found that using extractAPAAC could slightly improve the model prediction accuracy (increased to 0.1264 in Supplementary Table S7), after optimization of SVM parameters. Descriptor extractAPAAC extracted hydrophobicity, hydrophilicity and the sequence of amino acids from the sequence data.

The $R^2$ achieved is very low and that may be caused by three reasons. First, the descriptors used to transfer amino acid sequence into numerical values may not be suitable (the right information may not be extracted from sequence). Future work includes exploring more powerful descriptors or combing several descriptors to extract more features from sequence. The second reason is that the ML methods we used were not suitable, so we will explore others including convolutional neural network (CNN) and recurrent neural network (RNN). Last, there may be no association between amino acid sequence and protein yield at all. Until now, there is no report on predicting protein yield from its sequence. It is possible that this idea was explored but no correlation can be found between them. We need to use both biological and computational approaches to validate or falsify this hypothesis.

## Acknowledgements

We thank Wee Chin Wong for technical discussions and Shen Yifan for assistance.

## Funding

This work was supported by MOE Research Scholarship, MOE Tier-1 grant (R-279-000-452-133) and NRF CRP grant (R-279-000-512-281) in Singapore.

*Conflict of Interest:* none declared.